\documentclass{PoS}

\title{Lattice QCD analysis for instantaneous 
interquark potential in generalized Landau gauge}

\ShortTitle{instantaneous 
interquark potential in generalized Landau gauge}

\author{\speaker{Takumi Iritani}\\
        Kyoto University\\
        E-mail: \email{iritani@ruby.scphys.kyoto-u.ac.jp}}

\author{Hideo Suganuma\\
        Kyoto University\\
        E-mail: \email{suganuma@ruby.scphys.kyoto-u.ac.jp}}

\abstract{
  Using generalized Landau gauge, we study the continuous change of 
  gluon properties from the Landau gauge toward the Coulomb gauge 
  in SU(3) lattice QCD.
  We investigate ``instantaneous interquark potential'', which is defined 
  by the spatial correlation of the temporal link-variable $U_4$ 
  and is an interesting gauge-dependent concept.
  In the Coulomb gauge, the instantaneous potential is expressed 
  by the Coulomb plus linear potential, where the slope is, 
  however, 2-3 times as large as the physical string tension.
  In the Landau gauge, the instantaneous potential has no linear part.
  We find that the linear part is continuously growing 
  by varying gauge from the Landau gauge toward the Coulomb gauge.
  We also find that the instantaneous potential approximately reproduces 
  the physical interquark potential in a specific intermediate gauge,
  $\lambda_C$-gauge.
  This $\lambda_C$-gauge is expected to be a useful gauge 
  for modeling effective theories such as the quark potential model.
}
         
\FullConference{The XXVIII International Symposium on Lattice Field Theory, 
        Lattice2010\\
		June 14-19, 2010\\
		Villasimius, Italy}

\begin{document}

\section{Introduction}
  Quantum Chromodynamics (QCD) is the fundamental theory 
  of the strong interaction, and color SU(3) gauge symmetry is 
  one of the guiding principles to construct the theory. 
  In fact, QCD is formulated to satisfy the SU(3) gauge symmetry. 
  In actual calculations, however, one needs to choose some gauge 
  in order to remove redundant gauge degrees of freedom. 
  According to the choice of the gauge, different physical pictures 
  could be obtained, while physical quantities and phenomena 
  do not depend on the gauges.

  On the confinement, the most intuitive picture would be 
  the dual-superconductor effect, which was proposed by 
  Nambu, Mandelstam and 't Hooft \cite{NambuTHooftMandelstam}.
  In this scenario, the linear confinement potential between quarks 
  is derived from the one-dimensional formation of color-electric flux tube 
  in the dual-superconductor, and is mainly discussed 
  in the maximally Abelian gauge. 
  One of the other famous confinement scenarios is 
  Kugo-Ojima criterion \cite{KugoOjima}.
  In this scenario, the confinement is mathematically analyzed 
  in terms of the BRST charge, which is formulated in a covariant 
  and globally SU($N_c$) symmetric gauge such as the Landau gauge.

  In the Coulomb gauge, Gribov and Zwanziger proposed that 
  the confinement force is closely related to 
  ``instantaneous color Coulomb interaction''  
  between quarks \cite{Gribov,Zwanziger98}, 
  which is known as Gribov-Zwanziger scenario. 
  Greensite et al. showed that the instantaneous interaction 
  actually produces a linear interquark potential 
  from lattice QCD calculation \cite{Greensite0304}.
  However, the slope of the instantaneous potential is 2-3 times 
  larger than the actual value of the physical string tension.

  In this paper, we investigate the change of gluon properties 
  and physical picture by varying gauge continuously 
  from the Landau gauge toward the Coulomb gauge. 
  In particular, we focus on behavior of 
  the instantaneous interquark potential, 
  which is a key concept in Gribov-Zwanziger scenario.
  We also discuss about the linkage between QCD and
  the quark potential model, which is one of the most successful 
  effective models to describe hadron properties. 
 
\section{Formalism}
\subsection{Generalization of the Landau gauge}
  The Landau gauge is one of the most popular gauges in QCD, 
  and its gauge fixing is given by 
  \begin{equation}
    \label{eqLandaufix}
    \partial_\mu A_\mu = 0,
  \end{equation}
  where $A_\mu$ are $\mathrm{SU}(N_c)$ gauge fields.
  The Landau gauge keeps the Lorentz covariance 
  and global $\mathrm{SU}(N_c)$ color symmetry.
  In the Euclidean space-time, the Landau gauge is also defined as 
  the global condition to minimize the quantity
  $R_{\mathrm{Landau}} \equiv \int d^4 x \ \mathrm{Tr} 
    \left\{ A_\mu(x) A_\mu(x) \right\}$
  by gauge transformation \cite{Iritani}.

  The Coulomb gauge is also one of the most popular gauges, 
  and is defined as
  \begin{equation}
    \label{eqCoulombfix}
    \partial_i A_i = 0.
  \end{equation}
  This condition resembles the Landau gauge,
  but there are no constraints on $A_4$.
  In the Coulomb gauge, the Lorentz covariance is partially broken, 
  and gauge field components are completely decoupled into 
  $\vec{A}$ and $A_4$:
  $\vec{A}$ behave as canonical variables and 
  $A_4$ becomes an instantaneous potential.

  From Eqs.(\ref{eqLandaufix}) and (\ref{eqCoulombfix}),
  we consider generalization of the Landau gauge, which is defined as
  \begin{equation}
    \partial_i A_i + \lambda \partial_4 A_4 = 0.
  \end{equation}
  This generalized Landau gauge is called as ``$\lambda$-gauge'' \cite{Bernard}.
  By varying $\lambda$-parameter from $1$ to $0$, we can change the gauge 
  continuously from the Landau gauge toward the Coulomb gauge.

  The lattice QCD action is constructed from link-variables $U_\mu(x)
  \in \mathrm{SU}(N_c)$ 
  instead of the gauge fields $A_\mu(x) \in \mathfrak{su}(N_c)$, 
  and gauge fixing condition is also expressed in terms of $U_\mu(x)$.
  On the lattice, $\lambda$-gauge fixing is defined as the maximization of
  \begin{equation}
    R_\lambda[U] \equiv 
                 \sum_x \large\{ \sum_{i=1}^3 \mathrm{Re} \ \mathrm{Tr} \
    U_i(x) + \lambda \mathrm{Re} \ \mathrm{Tr} \ U_4(x) \large\}
  \end{equation}
  by gauge transformation of link-variables,
    $U_\mu(x) \rightarrow U'_\mu(x) = \Omega(x)U_\mu(x)\Omega^\dagger(x+\hat{\mu}),
    \ \Omega \in \mathrm{SU}(N_c)$.

\subsection{Terminated Polyakov line and instantaneous potential}
  After generalized Landau gauge fixing,
  we calculate ``$T$-length terminated Polyakov line'' $L(\vec{x},T)$, 
  which is defined as
  \begin{equation}
    L(\vec{x},T) \equiv U_4(\vec{x},1) U_4(\vec{x},2) \cdots U_4(\vec{x},T),
    \quad T = 1, 2, \dots, N_t
  \end{equation}
  on $N_s^3 \times N_t$ lattice.
  Here, we use the lattice unit of $a=1$. 
  We note that $L(\vec{x},T)$ is a gauge-dependent quantity.
  For $T = N_t$, the trace of $T$-length Polyakov line, 
  $\mathrm{Tr} \ L(\vec{x},N_t)$, results in the Polyakov loop.

  Using $T$-length Polyakov line,
  we define ``finite-time potential'' $V_\lambda(R,T)$ in $\lambda$-gauge as
  \begin{equation}
  \label{eqFTpotential}
    V_\lambda(R,T) \equiv - \frac{1}{T} 
    \ln \langle \mathrm{Tr} [L^\dagger(\vec{x},T)L(\vec{y},T)]\rangle, 
    \quad R = |\vec{x}-\vec{y}|.
  \end{equation}
  $V_\lambda(R,T)$ gives the energy between two color sources, 
  which are created at $t = 0$ and annihilated at $t = T$.

  Especially for $T = 1$, 
  we call $V_\lambda(R,1)$ as ``instantaneous potential''
  \begin{equation} \label{eqDefInstPot}
    V_\lambda(R) \equiv V_\lambda(R,1)
    = - \ln \langle
    \mathrm{Tr} [ 
      U_4^\dagger(\vec{x},1)U_4(\vec{y},1)
    ] \rangle, \qquad R = |\vec{x}-\vec{y}|.
  \end{equation}
  In the Coulomb gauge, the instantaneous potential
  is expressed by the Coulomb plus linear potential \cite{Greensite0304},
  while no linear part appears in this potential 
  in the Landau gauge \cite{Iritani,NakamuraSaito}.

\section{Lattice QCD calculation}
  We perform SU(3) lattice QCD Monte Carlo calculations on $16^4$ 
  with lattice-parameter $\beta = 5.8$ at the quenched level. 
  The lattice spacing $a$ is $0.152$fm, which is determined
  so as to reproduce the string tension as 
  $\sqrt{\sigma}_{\rm phys} = 427$MeV
  \cite{SuganumaTakahashiIchie}.
  We investigate the Landau gauge, the Coulomb gauge,
  and their intermediate gauges, i.e., $\lambda$-gauge with 
  $\lambda = 0.75, 0.50, 0.25, 0.10, 0.05, 0.04, 0.03$, $0.02, 0.01$.
  The number of gauge configurations is 50 for each $\lambda$.
  The statistical error is estimated by the jackknife method.

\subsection{Instantaneous potential}
\begin{figure}
  \begin{minipage}[t]{.47\textwidth}
    \centering
    \includegraphics[width=7cm,clip]
      {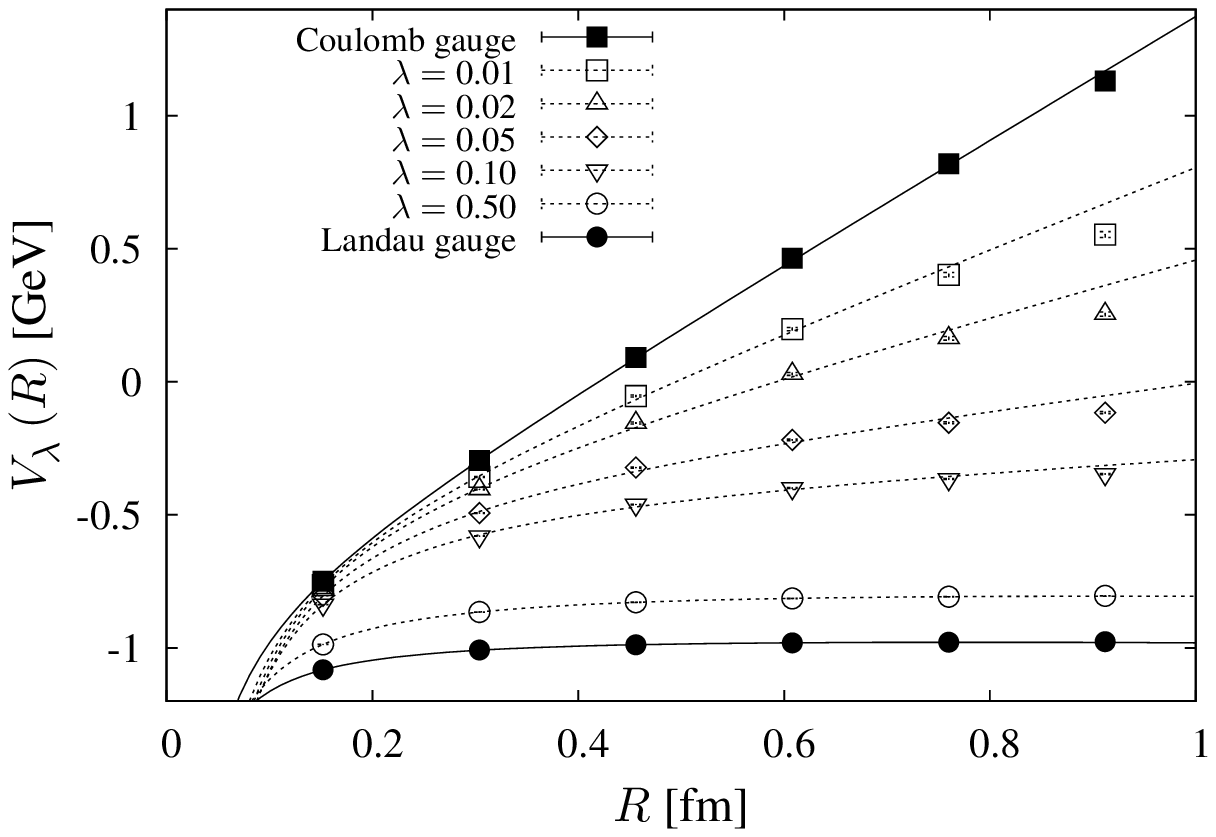}
    \caption{\label{figInstPot} 
      ``Instantaneous potential'' $V_\lambda(R)$ in generalized Landau gauge
      for typical values of $\lambda$.
      Symbols are lattice QCD results, and curves are
      fit results using Coulomb plus linear Ansatz.}
  \end{minipage}
  \hspace{0.5cm}
  \begin{minipage}[t]{.47\textwidth}
    \centering
    \includegraphics[width=7cm,clip]
      {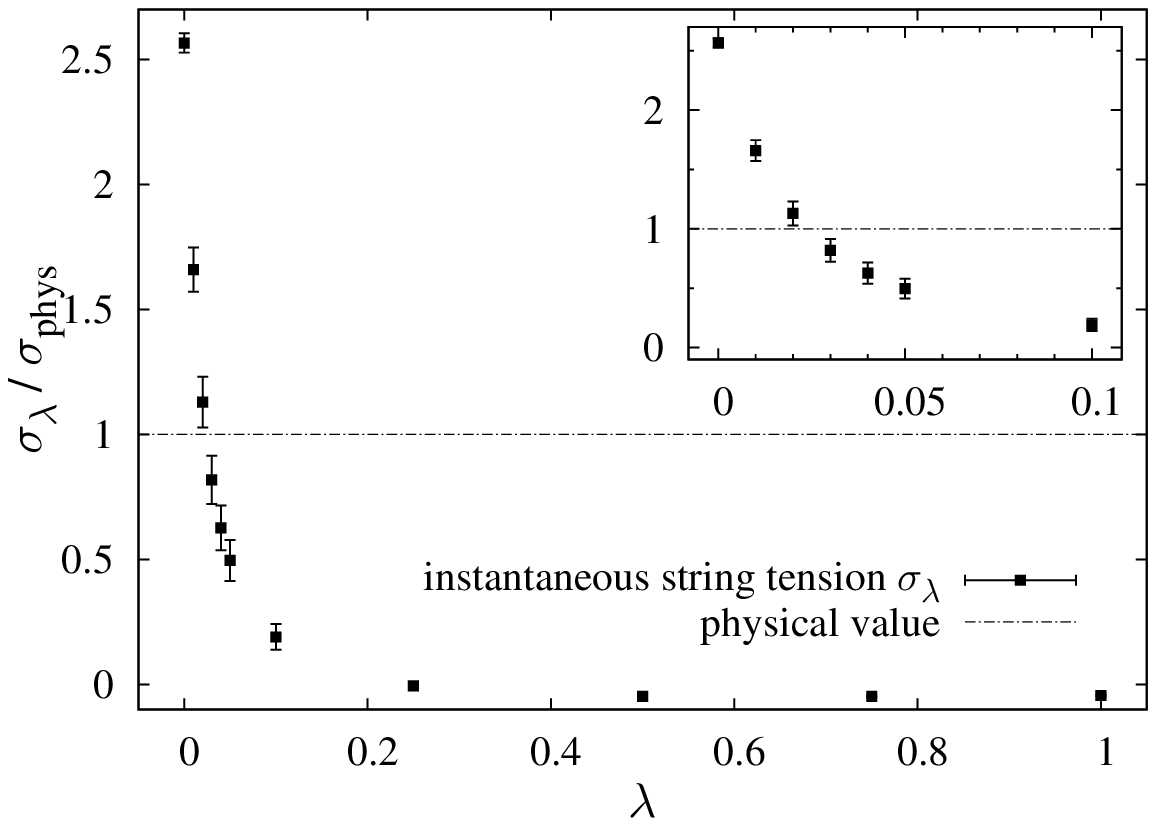}
    \caption{ \label{figStringTension}
      ``Instantaneous string tension'' $\sigma_\lambda$ 
      in generalized Landau gauge.
      $\sigma_\lambda$ changes continuously 
      from the Landau gauge to the Coulomb gauge.
      $\sigma_\lambda$ coincides with physical value $\sigma_{\rm phys}$ 
      at $\lambda_C \sim 0.02$.
      }
  \end{minipage}
\end{figure}

  We investigate the instantaneous potential $V_\lambda(R)$ 
  in generalized Landau gauge.
  Figure \ref{figInstPot} shows gauge dependence of $V_\lambda(R)$.
  In this figure, the statistic error is small and hidden in the symbols.
  
  In the Coulomb gauge ($\lambda = 0$), the instantaneous potential 
  shows linear behavior, while there is no linear part at all 
  in the Landau gauge ($\lambda = 1$).
  Thus, there is a large gap between these gauges 
  in terms of the instantaneous potential. 
  By varying gauge from the Landau gauge toward the Coulomb gauge,
  the potential $V_\lambda(R)$ grows monotonically, 
  and these two gauges are connected continuously. (See Fig.\ref{figInstPot}.)

  To analyze the instantaneous potential quantitatively,
  we fit the lattice QCD results using 
  Coulomb plus linear functional form as
  \begin{equation}
    V_\lambda(R) = - \frac{A_\lambda}{R} + \sigma_\lambda R + C_\lambda,
    \label{eqCoulplusLin}
  \end{equation}
  where $A_\lambda$ is Coulomb coefficient, $\sigma_\lambda$ slope
  of the potential (string tension), and $C_\lambda$  a constant.
  Here, besides the Coulomb plus linear Ansatz, 
  we try several candidates of the functional form,
  $-A/R + \sigma(1-e^{-\varepsilon R})/\varepsilon$, $-A\exp(-mR)/R$,
  $-A/R + \sigma R^d$, and $-A/R^d$, but they are less workable.
  The curves in Fig. \ref{figInstPot} are the best-fit results using
  Eq.(\ref{eqCoulplusLin}). 
  The Coulomb plus linear Ansatz works well 
  at least for $R \lesssim 0.8$fm, 
  which is relevant region for hadron physics.
  In the deep IR limit, $R \rightarrow \infty$, $V_\lambda(R)$ 
  goes to a saturated value, except for $\lambda = 0$.

  We focus on the gauge dependence of the linear slope $\sigma_\lambda$,
  which we call ``instantaneous string tension''. 
  (See Fig.\ref{figStringTension}.)
  For $\lambda \gtrsim 0.1$, $\sigma_\lambda$ is almost zero,
  so that this region can be regarded as ``Landau-like.''
  For $\lambda \lesssim 0.1$, 
  $V_\lambda(R)$ is drastically changed near the Coulomb gauge, 
  and $\sigma_\lambda$ grows rapidly in this small region.
  Finally, in the Coulomb gauge, one finds 
  $\sigma_\lambda \simeq 2.6\sigma_{\rm phys}$ 
  ($\sigma_{\rm phys} = 0.89$GeV/fm).
  Thus, the slope of the potential grows continuously from 
  the Landau gauge ($\sigma_\lambda \simeq 0$) towards the Coulomb gauge 
  ($\sigma_\lambda \simeq 2.6\sigma_{\rm phys}$), and therefore
  there exists some specific $\lambda$-parameter of $\lambda_C$ 
  where the slope of the instantaneous potential coincides 
  with the physical string tension.

\begin{figure}
  \begin{minipage}[t]{.47\textwidth}
    \centering
    \includegraphics[width=7cm,clip]
      {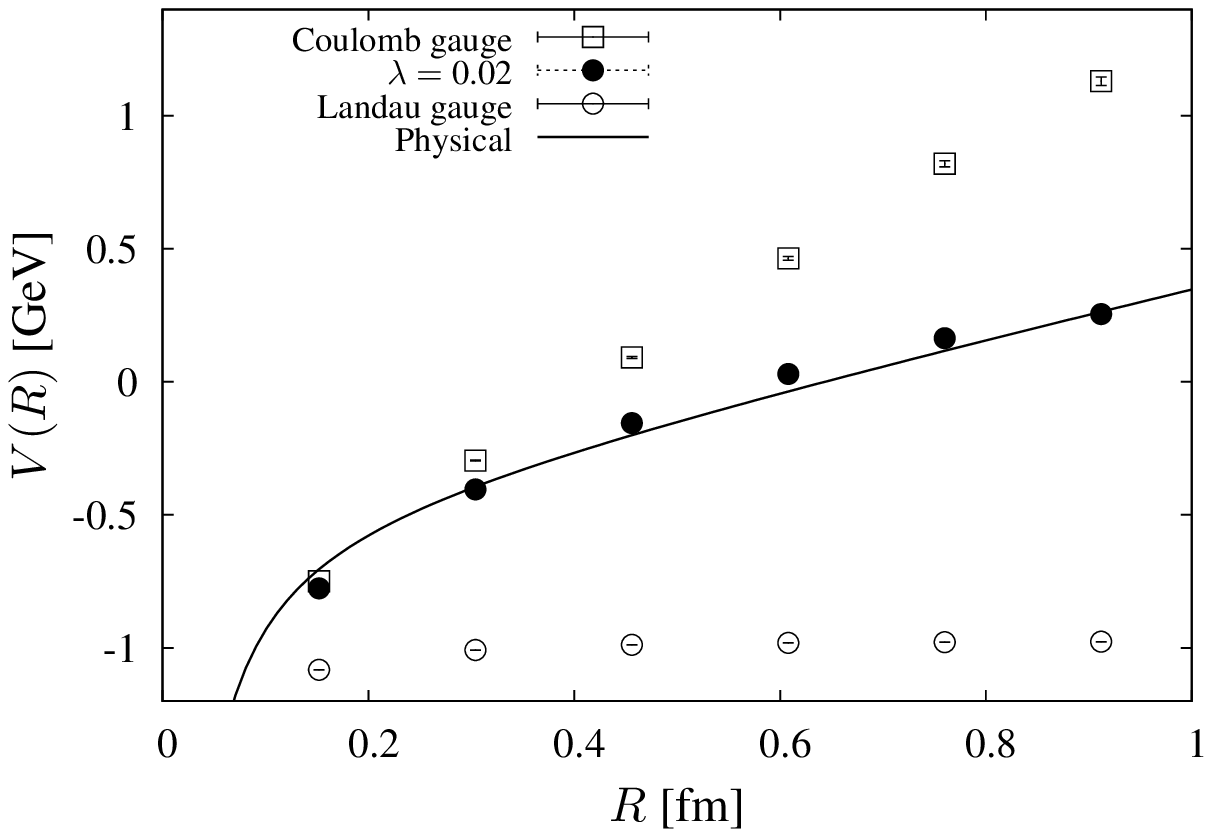}
    \caption{
      \label{figLambdaCPot}
      The instantaneous potential at $\lambda = 0.02$ $(\sim \lambda_C)$,
      the solid line is physical interquark potential 
      $V_{\rm phys}(R) = -A_{\rm phys}/R + \sigma_{\rm phys} R$ 
      with $A_{\rm phys} = 0.27$, and $\sigma_{\rm phys} = 0.89$GeV/fm.
    }
  \end{minipage}
  \hspace{0.5cm}
  \begin{minipage}[t]{.47\textwidth}
    \centering
    \includegraphics[width=7cm,clip]{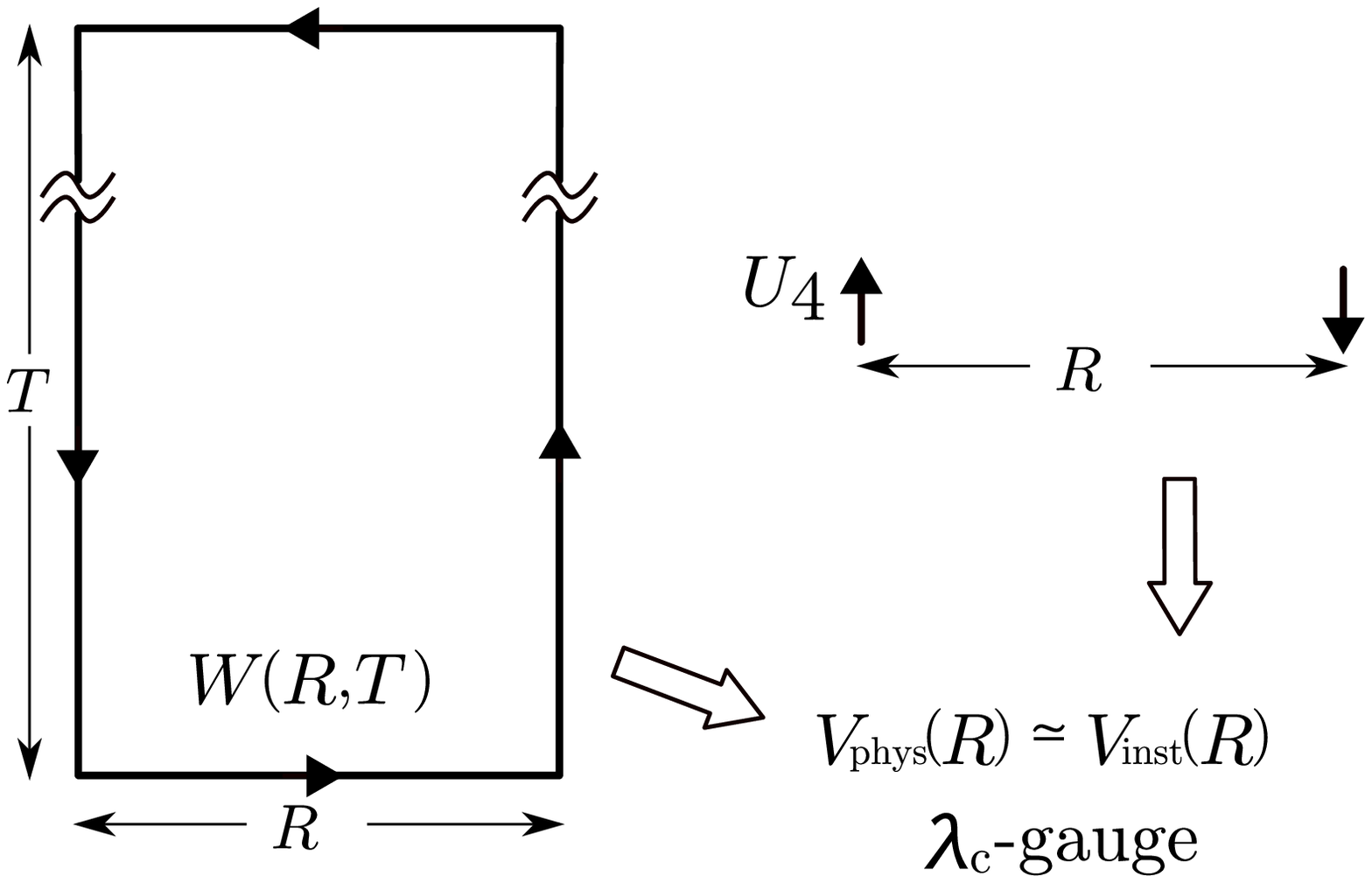}
    \caption{ 
      \label{figInstPotAndPhysPot}
      Schematic picture of physical interquark potential
      and instantaneous potential.
      In $\lambda_C$-gauge, instantaneous potential $V_{\rm inst}$
      approximately reproduces the physical potential $V_{\rm phys}$.
    }
  \end{minipage}
\end{figure}

  From Fig.\ref{figStringTension},
  the value of $\lambda_C$ is estimated to be about 0.02.
  In this $\lambda_C$-gauge, the physical static interquark potential 
  $V_{\rm phys}(R)$ is approximately reproduced by 
  the instantaneous potential. (See Fig.\ref{figLambdaCPot}.)
  While $V_{\rm phys}(R)$ is derived from large $T$
  behavior of the Wilson loop $W(R,T)$ as
  $V_{\rm phys}(R) 
  = - \lim_{T\rightarrow \infty} \frac{1}{T} \ln \langle W(R,T)\rangle$,
  only instantaneous correlation of $U_4$
  approximately reproduces the physical static potential in $\lambda_C$-gauge. 
  (See Fig.\ref{figInstPotAndPhysPot}.)

\subsection{Finite-time potential}
  Next, we analyze finite-time potential $V_\lambda(R,T)$ 
  defined by Eq.(\ref{eqFTpotential}), 
  which is a generalization of instantaneous potential.

\begin{figure}
  \begin{minipage}[t]{.47\textwidth}
    \centering
    \includegraphics[width=7cm,clip]
    {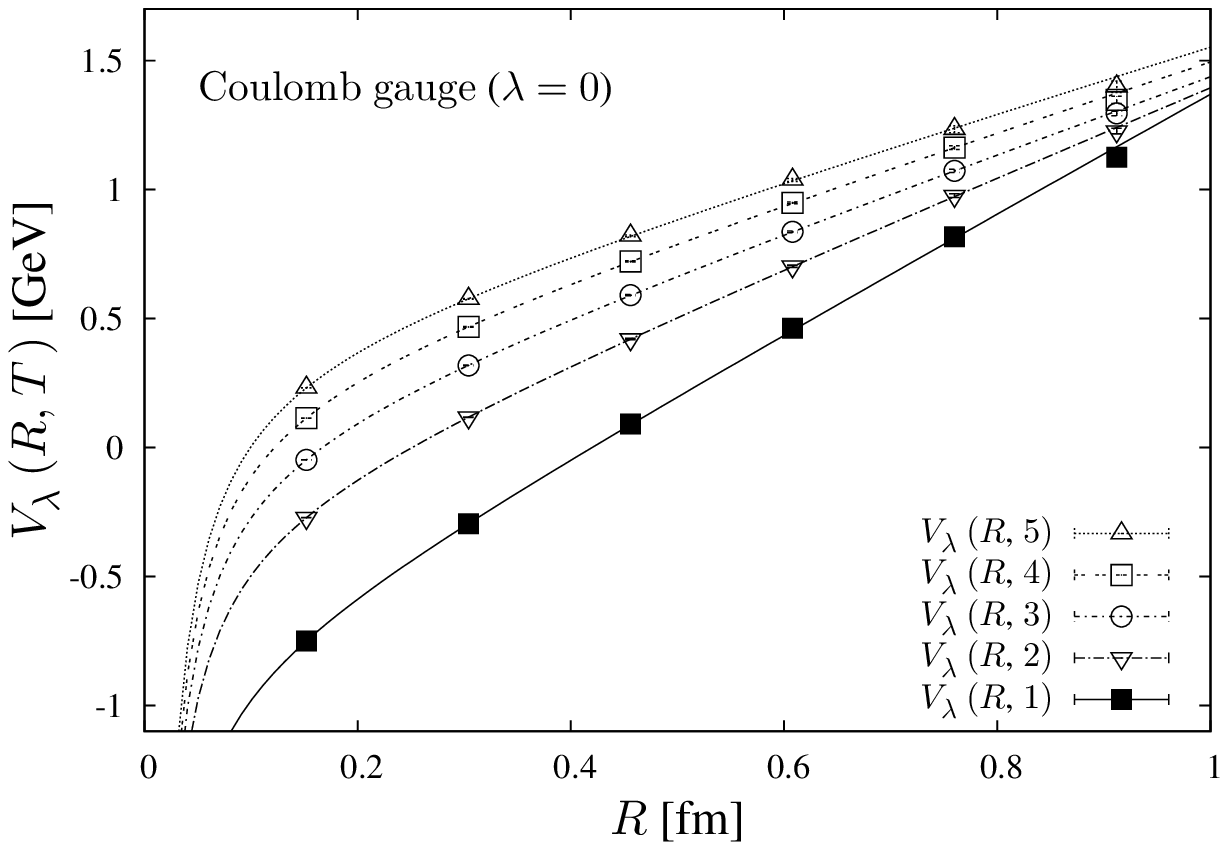}
    \caption{
    \label{figTlengthCoulomb}
    ``Finite-time potential'' $V_\lambda(R,T)$
    in the Coulomb gauge $(\lambda = 0)$.
    An irrelevant constant is shifted.
    The curves are the fit results using
    Coulomb plus linear function.
    }
  \end{minipage}
  \hspace{0.5cm}
  \begin{minipage}[t]{.47\textwidth}
    \centering
    \includegraphics[width=7cm,clip]
    {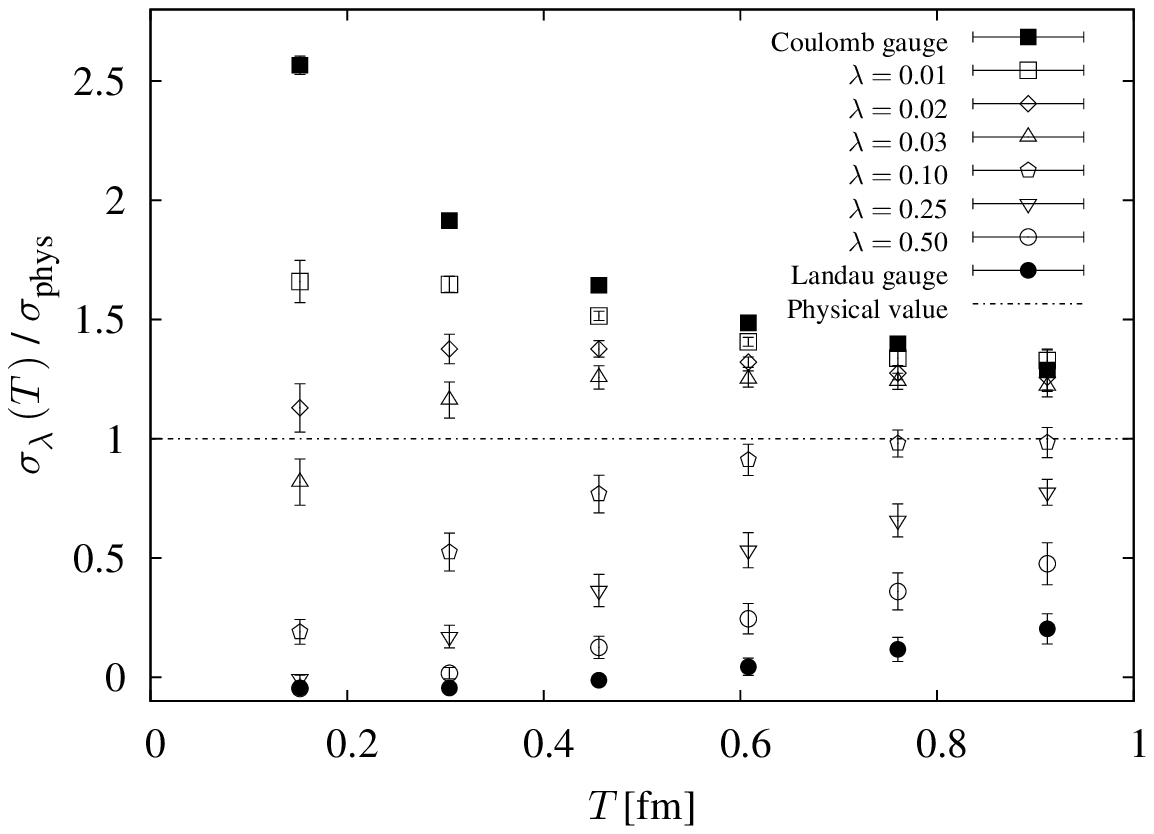}
    \caption{
    \label{figTlengthSlope}
    $T$-length dependence of the slope $\sigma_\lambda(T)$ of 
    finite-time potential $V_\lambda(R,T)$ 
    in generalized Landau gauge for several typical $\lambda$-values.
    }
  \end{minipage}
  \end{figure}

  First, we consider the Coulomb gauge.
  Figure \ref{figTlengthCoulomb} shows 
  the lattice QCD result for $V_\lambda(R,T)$ in the Coulomb gauge. 
  Similar to the instantaneous potential,
  $V_\lambda(R,T)$ is well reproduced by the Coulomb plus linear form.
  However, the parameter values are changed according to $T$-length.
  In particular, the slope of the potential becomes smaller 
  as $T$ becomes larger, which shows an ``instability'' of 
  $V_\lambda(R,T)$ in terms of $T$ in the Coulomb gauge.

  For general $\lambda$, 
  finite-time potential $V_\lambda(R,T)$ is found to be reproduced 
  by the Coulomb plus linear form as 
  \begin{equation}
    V_\lambda(R,T) 
    = - \frac{A_\lambda(T)}{R} + \sigma_\lambda(T) R + C_\lambda(T),
  \end{equation}
  at least for $R \lesssim 0.8$fm, similarly for the instantaneous potential. 

  We focus on $T$-length dependence of the slope $\sigma_\lambda(T)$ 
  of $V_\lambda(R,T)$ at each $\lambda$. (See Fig.~\ref{figTlengthSlope}.) 
  In the Coulomb gauge ($\lambda=0$), $\sigma_\lambda(T)$ is 
  a decreasing function: starting from 2-3 times larger value, 
  it approaches to the physical string tension $\sigma_{\rm phys}$, 
  as $T$ increases.
  Around $\lambda_C$-gauge, i.e., for $\lambda \sim \lambda_C(\simeq 0.02)$, 
  $T$-dependence is relatively weak, and $\sigma_\lambda(T)$ seems to 
  converge on the same value of about $1.3 \sigma_{\rm phys}$ 
  around $T \sim 1$fm. 
  For $\lambda \gtrsim 0.1$ (Landau-like), 
  $\sigma_\lambda(T)$ is an increasing function of $T$: 
  starting from zero at $T=1$, 
  the linear part of $V_\lambda(R,T)$ appears and grows, 
  as $T$ increases.
  At each $\lambda$, $\sigma_\lambda(T)$ seems to approach to 
  the physical string tension $\sigma_{\rm phys}$ for sufficiently large $T$.

\section{Summary and Discussion}
  In this paper, aiming to grasp the gauge dependence of gluon properties,
  we have investigated generalized Landau gauge and applied it to
  instantaneous interquark potential in SU(3) lattice QCD at $\beta$=5.8.
  In the Coulomb gauge, the instantaneous potential is expressed by 
  the sum of Coulomb potential and linear potential 
  with 2-3 times larger string tension. 
  In contrast, the instantaneous potential has no linear part
  in the Landau gauge. Thus, there is a large gap between these two gauges.
  Using generalized Landau gauge, we have found that the instantaneous 
  potential $V_\lambda(R)$ is connected continuously from the Landau gauge 
  towards the Coulomb gauge, and the linear part in $V_\lambda(R)$ 
  grows rapidly in the neighborhood of the Coulomb gauge.
  
  Since the slope $\sigma_\lambda$ of the instantaneous potential 
  $V_\lambda(R)$ grows continuously from 0 to 2-3$\sigma_{\rm phys}$, 
  there must exist some specific intermediate gauge 
  where the slope $\sigma_\lambda$ coincides with 
  the physical string tension $\sigma_{\rm phys}$.
  From the lattice QCD calculation, the specific $\lambda$-parameter,
  $\lambda_C$, is estimated to be about $0.02$.
  In this $\lambda_C$-gauge, the physical static interquark potential 
  $V_{\rm phys}(R)$ is approximately reproduced 
  by the instantaneous potential $V_\lambda(R)$. 
  (See Fig.\ref{figInstPotAndPhysPot}.)

  We have also investigated finite-time potential $V_\lambda(R,T)$, which 
  is defined from $T$-length terminated Polyakov line and a generalization
  of the instantaneous potential.
  The behavior of the slope $\sigma_\lambda(T)$ of finite-time potential 
  is classified into three groups: the Coulomb-like gauge 
  ($\lambda \lesssim 0.01$), the Landau-like gauge ($\lambda \gtrsim 0.1$), 
  and neighborhood of $\lambda_C$-gauge ($\lambda \sim \lambda_C$).
  In the Coulomb-like gauge, the slope $\sigma_\lambda(T)$ 
  is a decreasing function of $T$, and seems to approach to 
  physical string tension $\sigma_{\rm phys}$ for large $T$.
  In the Landau-like gauge, $\sigma_\lambda(T)$ is an increasing function.
  Around the $\lambda_C$-gauge,
  $\sigma_\lambda(T)$ has a weak $T$-length dependence.

  Finally, we consider a possible gauge of QCD to describe
  the quark potential model from the viewpoint of instantaneous potential.
  The quark potential model is a successful nonrelativistic framework 
  with a potential instantaneously acting among quarks, 
  and describes many hadron properties in terms of quark degrees of freedom.
  In this model, there are no dynamical gluons, and gluonic effects 
  indirectly appear as the instantaneous interquark potential.

  As for the Coulomb gauge, the instantaneous potential has too large 
  linear part, which gives an upper bound on the static potential 
  \cite{Zwanziger03}. It has been suggested by Greensite et al. that
  the energy of the overconfining state is lowered by inserting 
  dynamical gluons between (anti-)quarks,
  which is called ``gluon-chain picture''.
  This gluon-chain state is considered as the ground state 
  in the Coulomb gauge \cite{Greensite0304,GreensiteThorn}.
  Therefore, dynamical gluon degrees of freedom must be 
  also important to describe hadron states in the Coulomb gauge.

  For $\lambda_C$-gauge, the physical interquark potential 
  $V_{\rm phys}(R)$ is approximately reproduced 
  by the instantaneous potential $V_{\lambda_C}(R)$. 
  This physically means that all other complicated effects
  including dynamical gluons and ghosts are approximately cancelled
  in the $\lambda_C$-gauge, and therefore we do not need to 
  introduce any redundant gluonic degrees of freedom.
  The absence of dynamical gluon degrees of freedom
  would be a desired property for the quark model picture.
  The weak $T$-length dependence of $\sigma_\lambda(T)$
  around the $\lambda_C$-gauge ($T$-length stability)
  is also a suitable feature for the potential model.
  In this way, as an interesting possibility,
  the $\lambda_C$-gauge is expected to be a useful gauge
  in considering the linkage from QCD to the quark potential model.

\section*{Acknowledgements}
  This work is supported by the Global COE Program, 
  ``The Next Generation of Physics, Spun from Universality
  and Emergence'' at Kyoto University.
  H.S. is supported in part by the Grant for Scientific Research
  [(C) No. 19540287, Priority Areas ``New Hadrons'' (E01:21105006)]
  from the Ministry of Education, Culture, Science, and Technology (MEXT) 
  of Japan.
  The lattice QCD calculations have been done on NEC-SX8 at Osaka University.

\end{document}